# Generation of highly reactive oxygen species by co-adsorption of oxygen and water on metal-supported MgO(100) thinfilms

Zhenjun Song, Hu Xu*

Department of Physics, South University of Science and Technology of China, Shenzhen, 518055, China

*Supporting Information Placeholder*

**ABSTRACT:** The formation of highly reactive oxygen species (ROS) on metal oxide surfaces have attracted considerable interest due to their diverse applications. In this work, we have performed density-functional theory calculations to investigate the co-adsorption of oxygen and water on ultrathin MgO(100) films deposited on Mo(100) substrate. We reveal that the molecular oxygen can be stepwise decomposed completely with the assistance of water. Consequently, a series of highly ROS including superoxide, hydroperoxide, hydroxyl and single oxygen adatom are formed on Mo(100) supported MgO(100) thinfilms. The reaction barriers accompanied by the generation of ROS are reported, and the influence of the thickness of MgO(100) films is also discussed. The most promising routes to produce these fascinating species provide valuable information to understand the importance of synergistic effect, namely the substrate, the co-adorbed species, and the film thickness, in multiphase catalyst design.

Reactive oxygen species (ROS) play essential roles in chemical and biological processes. Understanding the mechanisms of how to generate ROS on metal oxide surfaces is of fundamental interest, as we can selectively control the chemical reactions if we know how to generate ROS. Recently, there has been a surge in researches pertaining to the physical and chemical properties of multiphase interfaces.[1-2] These interactions between solid, liquid, and gas play a key role in solv-ing many kinds of scientific questions, such as environmental sciences, bioengineering, microelectronics, etc. In addition, the strategic interaction among reactants and substrate is also beneficial for achieving efficient catalysis of multiphase reactions, because of the large interfacial contact areas.[3]

Water is the most important liquid and the most frequently used solvent, and molecular oxygen is the most common gas with strong oxidizing property. Can we design a comprehensive research model to study the multiphase catalytic reactions? The co-adsorption of atomic oxygen and water has been intensively studied on Ru(0001) surface.[2,3] Weinberg et al.[2] showed that water does not react with preadsorbed oxygen to form OH group, while Held et al.[3] pointed out that water can only form partial dissociation configuration for oxygen coverages below 0.18 monolayer (ML) at 140 K. In addition, the reaction between $H_2O$ and atomic O to form OH group is also reversible,[3-7] and all OH groups will recombine to form water and desorb after annealing to 200 K. Moreover, the interaction of molecular oxygen and water has been widely studied on reduced metal oxide surfaces. The intact water molecules are energetically most favorable, and hydroxyl groups are unstable.[4-6] However, till now it remains challenging and desirable to generate ROS on insulating surfaces. Therefore, how to produce the highly ROS in a practical avenue and inexpensively, is a meaningful subject worth being investigated.

The pure stoichiometric MgO(100) is one of typical insulating surfaces, which lacks intrinsic catalytic activity. However, the MgO(100) surface combining with transition metal clusters or substrates presents an outstanding catalytic effect as well as novel electronic properties,[7-9] and the growing behavior and final morphology of ultrathin MgO films can even be tuned by transition metal substrate.[10-11] As it is well known, electron transfer between ultrathin insulating films and transition metal substrates plays crucial roles in many chemical process, which is frequently employed toward technically relevant applications and heterogeneous catalysis.[12] The first observation of electron transfer on MgO surface has been confirmed by Lunsford and Jayne in 1978.[13] Recently, Gonchar et al[14] provide experimental evidence for the formation of superoxide radicals on MgO(100)/Mo(100) surface by electron paramagnetic resonance (EPR) spectroscopy. This activated molecular oxygen species, due to electron transfer mechanism, is important in understanding the excellent catalytic properties for CO oxidation.[15] In this study, we have examined systemically the co-adsorption of water and oxygen molecules on Mo(100) supported ultrathin MgO(100) films. As water and molecular oxygen always present in surface photocatalytic experiments under ambi-ent conditions, therefore the study of the interaction between water and molecular oxygen is of significant importance for understanding and enhancing the catalytic performance of multiphase reactions. More importantly, we have obtained a series of highly ROS on supported ultrathin MgO films, with very low energy barrier and very high heat release.

Density functional theory (DFT) calculations with spin polarization are performed using the Vienna Ab Initio Simulation (VASP) package.[16-18] Projector augmented wave (PAW) method,[19-21] and Perdew, Burke, and Ernzerhof (PBE) functional[22-23] with generalized gradient approximation (GGA) are used. The energy cutoff is 500 eV, and the successive slabs are separated by a vacuum distance larger than 17 Å. The substrate is built using four atomic Mo layers, and the bottom two Mo atomic layers are frozen at their equilibrium bulk positions to mimic bulk properties. One to five ML of MgO(100) are adopted as ultrathin MgO films. The p(4x4) Mo(100) substrate is employed in the calculations, and there are 16 O and 16 Mg atoms per layer. Each atom is relaxed until the residual force acting on each atom is less than 0.02 eV/Å, while a (2×2×1) k-point sampling is used for structural relaxations. Dipole corrections, which are tested on selected cases, do not result in different geometric structures and adsorption properties. The reaction pathways and barriers are calculated using the climbing image nudged elastic band (CI-NEB) method.[24]

It is generally known that the initial adsorption behavior of $O_2$ and water on the catalyst surface affects the subsequent reactions significantly. $O_2$ will only be weakly physisorbed on stoichiometric MgO(100) surface, and the binding energy per $O_2$ is only 24 meV. By contrast, $O_2$ interacts strongly with Mo(100) supported ultrathin MgO(100) films, and $O_2$ prefers to locate symmetrically between the two neighboring Mg ions,[14] as shown in Fig. 1a. The respective bond lengths of the optimized $O_2$ dimers on 1-5 ML MgO/Mo(100) surfaces are 1.405 Å, 1.380 Å, 1.360, 1.353 Å, and 1.346 Å, which are much longer than that of gaseous $O_2$ of 1.233 Å. From Bader charge analysis, we can find that $O_2$ dimer on Mo-supported ultrathin MgO films is negatively charged, and the charge gains are 1.16 electron (*e*), 1.10 *e*, 0.99 *e*, 0.88 *e*, and 0.84 *e*, respectively. The results indicate that molecular oxygen is activated and becomes superoxide species, which agrees well with the experimental results.[14] The adsorption energies per $O_2$ for 1-5 ML MgO/Mo(100) surfaces are -2.08 eV, -1.88 eV, -1.59 eV, -1.44 eV and -1.29 eV, respectively. Our results indicate that the adsorption energy and bond length of $O_2$ closely depend on the amount of charge transfer between $O_2$ and MgO(100)/Mo(100), and the charge transfer decreases with the increasing the thickness of MgO films. As a result, the electronic state of $O_2$ is changed after adsorption, and the molecular mag-netic moment of $O_2$ is greatly reduced along with the decrease of the thickness of MgO film (0.60 $\mu_B$, 0.82 $\mu_B$, 0.95 $\mu_B$, 1.00 $\mu_B$, and 1.05 $\mu_B$ for $O_2$ on the 1-5 ML MgO/Mo(100) and 1.96 $\mu_B$ for $O_2$ on the bulk MgO(100)). On the pure MgO(100) film, the O-O bond length of the coadsorbed $O_2$ is 1.247 Å, and Bader charge of $O_2$ is -0.11 $e$. Our results indicate that both the superoxide and hydroperoxide species cannot form on pure MgO(100). Thus, the structure and property of oxygen molecule remain almost unchanged on the pure MgO(100). The results reveal that metal substrate plays a very important role in activating oxygen molecule, although the metal substrate does not serve as the active sites and there is no direct contact between $O_2$ and the metal substrate.

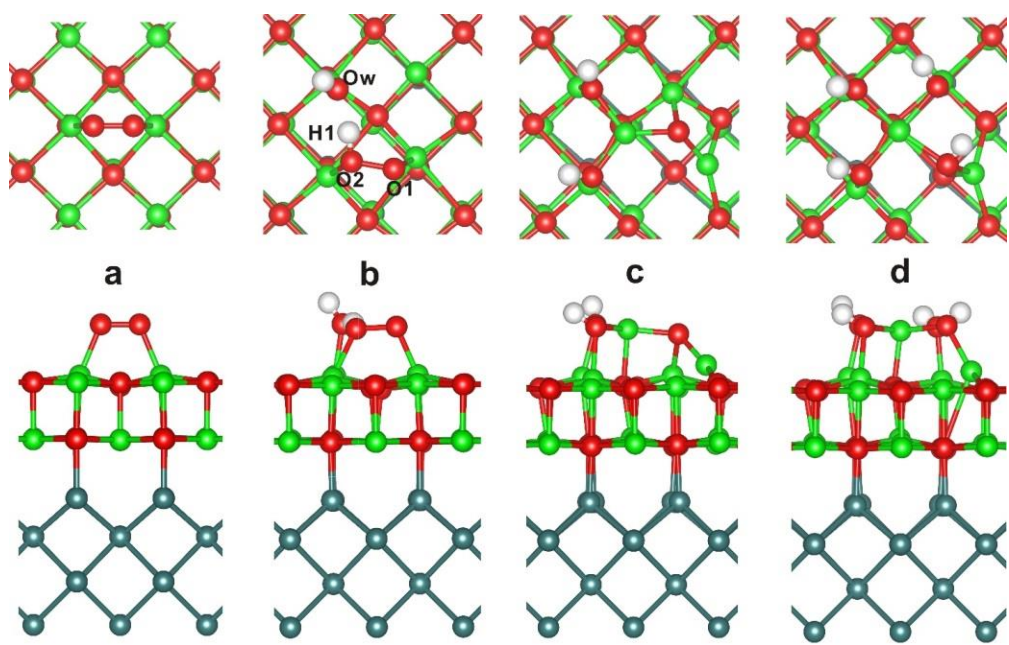

**Figure 1.** Top and side views of optimized geometries of $O_2$ or/and water on the 2 ML MgO/Mo(100) surface. (a) $O_2$ adsorption, (b) $O_2$ and water forming OOH and OH, (c) $O_2$ and water forming 2OH and an isolated oxygen adatom, and (d) $O_2$ and two water molecules forming four OH groups.

The interaction of water with Mo supported MgO films has been studied by us recently.[25] Our results show that water prefers to dissociate on ultrathin MgO(100) films induced by in-plane tensile strain between Mo(100) substrate and ultrathin MgO films. Next, we have studied the co-adsorption of water and $O_2$ on MgO(100)/Mo(100). When water meets $O_2$ on the 1-3 ML MgO/Mo(100) surfaces , the hydrogen of water is snatched by $O_2$, and then the OOH species forms. The optimized geometry and structural parameters are shown in Figure 1b and Table 1, respectively. The Bader charge of both O1 and O2 of OOH on 1-3 ML MgO/Mo(100) surfaces is in the range from -1.43 $e$ to −1.54 $e$, which deviates from peroxide of $H_2O_2$ (-1.17 $e$) a lot due to their different bonding environment, although they have the similar bond-length. The peroxide in $H_2O_2$ show strong covalent bonding between oxygen and hydrogen, while the O1O2 of OOH group on the MgO/Mo(100) present not only covalent bonding with H1, but also ionic bonding with fivefold coordinated surface Mg atoms. Because of the work function reduction of the Mo(100),[26] partial charge will transfer from the substrate to the adsorbed species with high electron affinity, resulting in the strong ionic bonding between OOH and MgO thinfilms . Thus, the OO of hydrogenated superoxide OOH on MgO/Mo(100) is not superoxide group any more. The much larger amount of charge of O1O2 in OOH group than $O_2$ adsorbed on MgO/Mo(001) and the bond length extremely close to O-O of $H_2O_2$, which prove that the OOH is an negatively charged hydroperoxide species. The calculated Bader charge of OOH on 1 ML, 2 ML and 3 ML ultrathin MgO films are -0.88 $e$, -0.90 $e$ and -0.80 $e$, respectively, which reconfirm that the superoxide radical anion is definitely reduced to peroxide species. However, OOH species does not form when the thickness of MgO films exceeds 4 ML. For thick MgO(100) films (n≥4), it is not possible for $O_2$ to capture enough electron to form OOH by exfoliating hydrogen from water.

The adsorption of oxygen molecule is exothermic greatly over Mo(100) supported ultrathin MgO(100) films, with adsorption energy of -2.08 ~ -1.59 eV eV for adsorption process on 1-3 ML films. The reaction pathways for the formation of OOH species are depicted in Figure 2. In the initial stage of reaction, water is far away from the surface, while $O_2$ adsorbs on the MgO/Mo(100) surface as the reactant. Surprisingly, the formation of OOH group is barrierless by the interaction of water and $O_2$. Meanwhile, the highly reactive OH group also forms on the surface. The barrierless reaction strongly indicates that hydroperoxide and hydroxyl groups will spontaneously form when water meets $O_2$ on 1-3 ML MgO/Mo(100) surfaces. This reaction is also exothermic. The reaction heat released for co-adsorption of water and $O_2$ on 1 ML and 2 ML MgO(100)/Mo(100) surfaces are respective 1.534 eV and 1.445 eV, while this value decreases dramatically to 0.822 eV on 3 ML MgO(100)/Mo(100) surface. Due to less reactive feature of thicker MgO(100) films (≥ 4ML), OOH group will not form on thicker MgO(100) films anymore.

Then we have considered the possible decomposition pathway of OOH group, and the corresponding reaction pathway are shown in Figure 3. The calculated energy gains on 1 ML, 2 ML and 3 ML MgO/Mo(100) surfaces are 4.768 eV, 3.617 eV and 1.388 eV, respectively, which clearly show that splitting OOH group into OH group and single oxygen adatom can dramatically lower the system energy. For 1 ML MgO(100)/Mo(100) surface, it needs to overcome an energy barrier of 0.633 eV in the reaction process, while the energy barrier is slightly increased to 0.776 eV for 2 ML MgO(100)/Mo(100) surface. These two energy barriers are relatively small, and it should be easy for OOH group to dissociate on supported ultrathin MgO(100) films(1ML ~ 2ML) at moderate temperature. We have also noted that the energy gain is only 1.388 eV for 3 ML MgO(100)/Mo(100) surface, and the reaction barrier height also increase to 1.084 eV. In other words, the ability to completely split molecular oxygen decreases with the increasing of the thickness of oxide films. Therefore, we have successfully proposed a novel catalyst system to decompose oxygen molecule on metal supported ultrathin MgO(100) films by introducing the water molecule to the reactants.

**Table 1. Structural parameters (unit in Å) and Bader charges $Q$ (unit in $e$) for co-adsorption of $O_2$ and water on *1-5* ML MgO(100)/Mo(100) surfaces and bulk MgO(100) surface.**

|  | $d_{O1-O2}$ | $d_{O1-H1}$ | $d_{Ow\cdots H1}$ | $Q_{O1+O2}$ | $Q_{O1+O2+H1}$ |
|---|---|---|---|---|---|
| 1 ML | 1.494 | 1.049 | 1.514 | -1.51 | -1.16 |
| 2 ML | 1.502 | 1.042 | 1.561 | -1.54 | -1.10 |
| 3 ML | 1.480 | 1.061 | 1.492 | -1.43 | -0.99 |
| 4 ML | 1.410 | 1.513 | 1.051 | -1.08 | -0.88 |
| 5 ML | 1.368 | 1.681 | 0.971 | -0.94 | -0.84 |
| bulk | 1.247 | 2.130 | 0.979 |  |  |

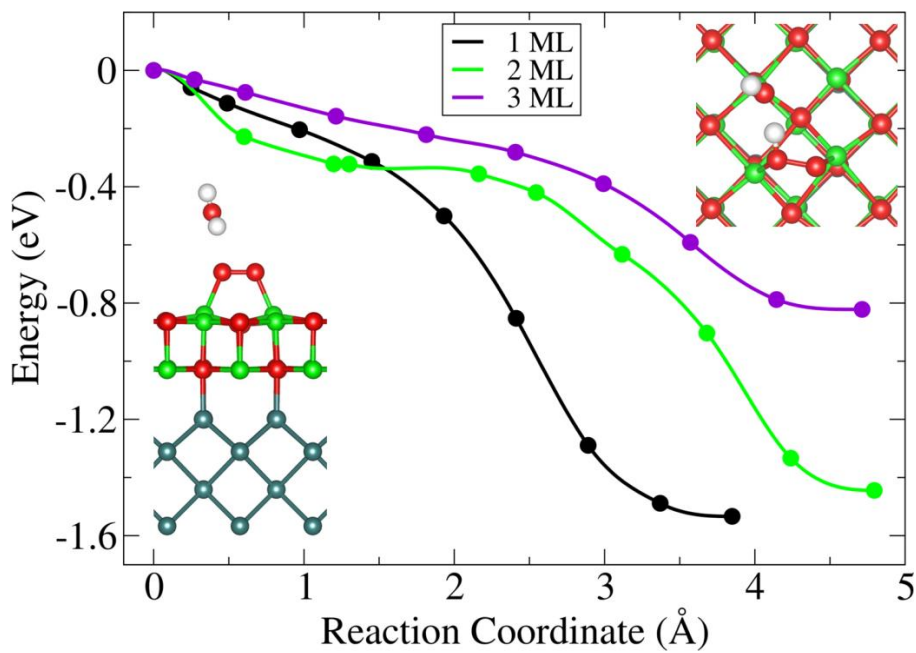

**Figure 2.** Potential energy profiles for hydroperoxide and hydroxyl species formation on 1-3 ML MgO/Mo(100) surfaces.

To further study the role of water in the generation of ROS, we have introduced another water molecule. The second water molecule lands on the surface and reacts with the oxygen adatom. This reaction leads to the formation of four hydroxyl groups, as shown in Fig. 1d. The interaction between oxygen adatom and water to form hydroxyl groups is barrierless (see Figure S1), and the associated energy gains are -2.10 eV, -2.18 eV, and -2.02 eV, respectively, for reaction occurring on 1-3 ML films. Thus, after the migration of oxygen atom of OOH group to neighboring surface site, the very reactive oxygen adatom can react with another water molecule easily, and finally all the oxygen and water are completely transformed to hydroxyl groups.

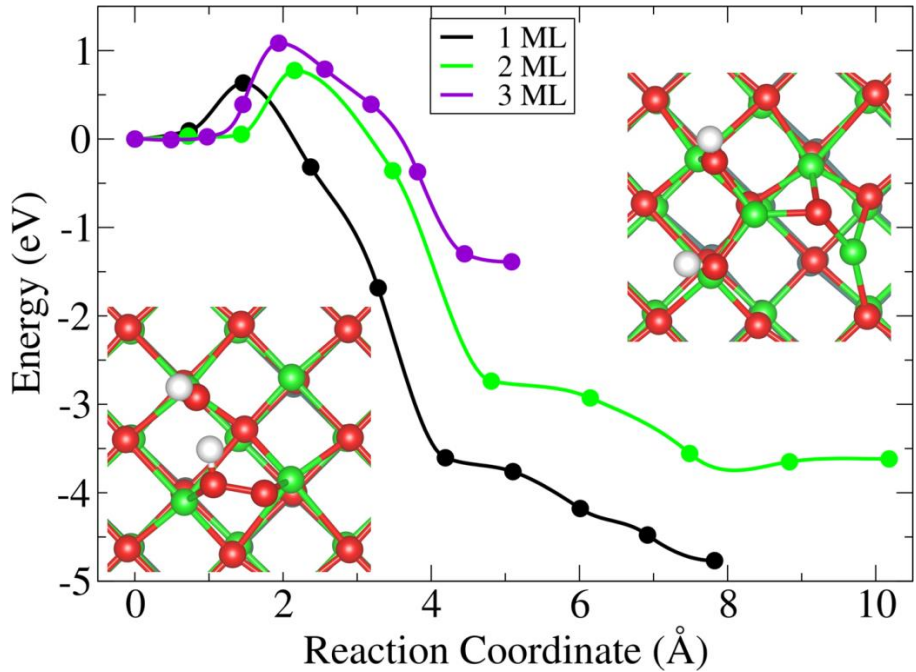

**Figure 3.** Potential energy profiles for producing high reactive oxygen adatom and hydroxyl species on 1-3 ML MgO/Mo(100) surfaces.

In the following, we turn to analyze electronic properties of MgO(100)/Mo(100) surfaces. When ultrathin MgO(100) films deposited on Mo(100) substrate, all the Mo atoms at interfacial region transfer electrons to oxygen, and the charge density changes demonstrate that the relatively strong chemical bonds form between O and Mo. On account of the high electron affinity of Mo, Mo atoms attract electrons from ultrathin MgO(100) films except the interface Mo atoms (Figure 5c). For $O_2$ adsorption on MgO(100)/Mo(100) surfaces, the electronic state distribution of the two oxygen atoms is identical, because of the same coordination environment and symmetric geometry. The adsorbed $O_2$ presents discrete state peaks, as depicted in Figure 4a, and the electron distribution near the Fermi level is mostly composed of $2p_z$ states of $O_2$.

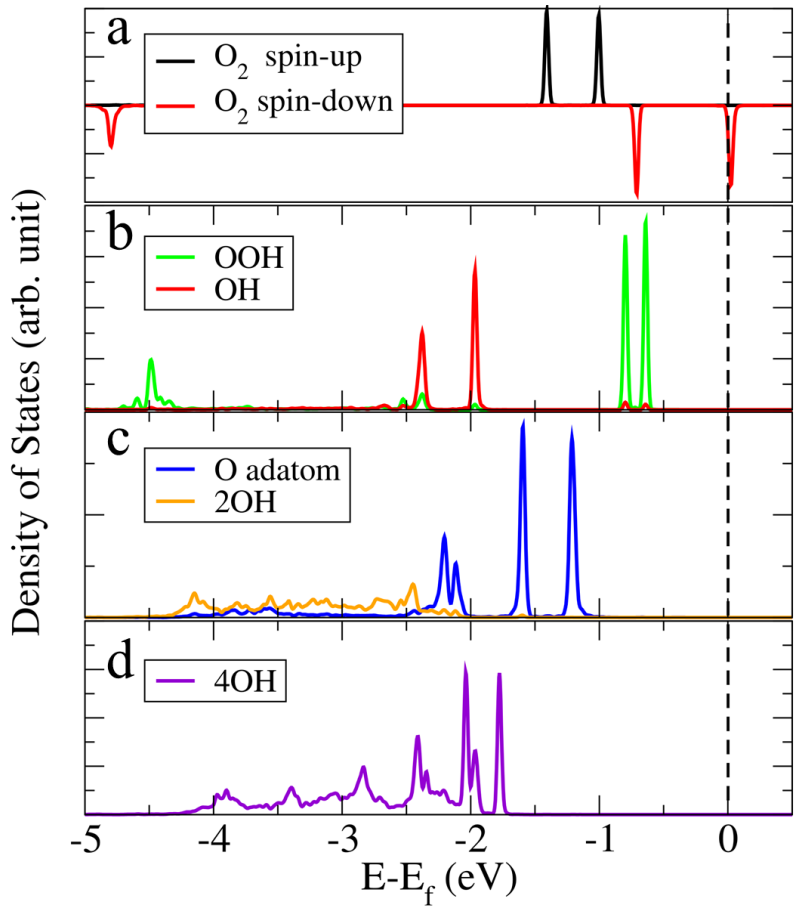

**Figure 4.** Density of states of O₂ and water on the 2 ML MgO/Mo(1001) surface. (a) The adsorbed O₂ adsorbed, (b) OOH and OH of dissociated water, (c) oxygen adatom and OH groups, and (d) fully dissociated OH groups.

As we have pointed that after we introduce one water molecule to react with $O_2$, OOH and OH groups can spontaneously form on Mo(100) supported 1-3 ML MgO(100) films, and these species on 1 ML and 2 ML MgO(100) films are spin unpolarized (Figure 4b). Through analysis of partial charge density distribution near Fermi level as depicted in Figure 5a and Figure 5b, we can find the electrons in metal supported MgO thinfilms enter the vicinity of Fermi level, and this metallic characteristic of MgO(100) films has crucial influence in surface reactions. In addition, from the differential charge density shown in Figure 5c, we demonstrate that OOH and OH groups prefer to withdraw electrons (green regions) from metal supported MgO films. We also find many blue regions (electron depletion) on the periphery of the green regions (electron accumulation) surrounding the OOH and OH species, indicating the formation of radicals with very high electron affinities on the metal supported ultrathin oxide films. $O_wH$ interacts with the surface strongly as its electronic states are widened extensively, and the electronic states are shift-ed to lower energy levels, especially for 1 ML and 2 ML films shown in Figure 4b. The strong interaction of $O_w$ from water and surface Mg contributes to stabilize the usually high reactive hydroxyl radical. While the electronic states of OOH group are shifted to higher energy region comparing with OH, indicating the strong chemical reactivity of OOH species. The OOH species on 3 ML film is slightly spin polarized, and the majority of spin polarization from the adsorbed $O_2$ keep almost unchanged. The spin polarization of OOH on slight thick MgO(100) films indicate that it is less favorable for $O_2$ to extract electrons. Moreover, $O_2$ and water do not react with each other on 4 ML MgO/Mo(100) film (Figure 5d), and therefore we can find π type spin density difference mostly concentrated on $O_2$.

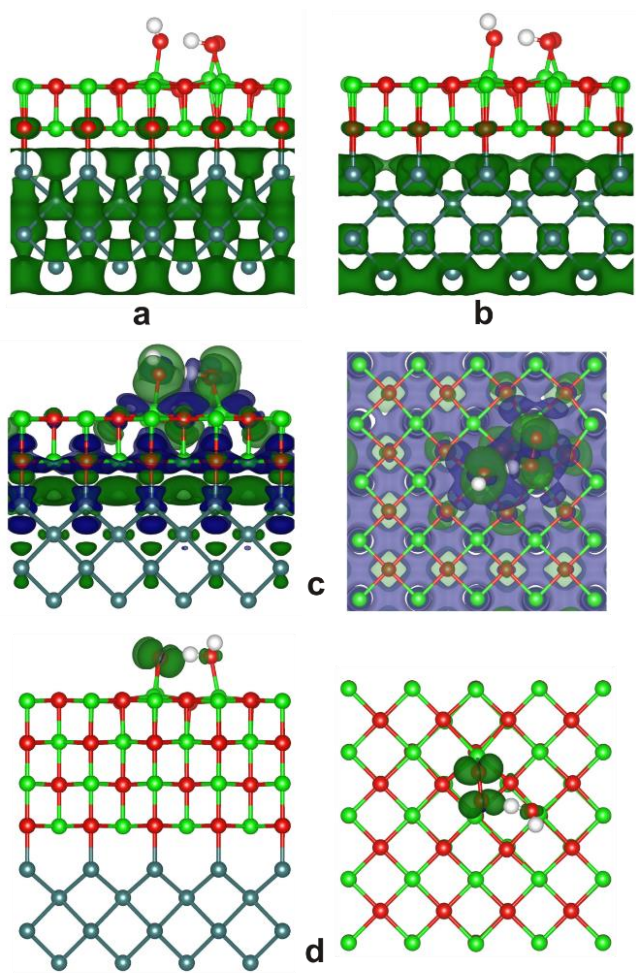

**Figure 5.** Charge density for the co-adsorption of $O_2$ and $H_2O$ on MgO/Mo(100) surface. (a) Partial charge density in the energy range of $E_f \sim E_f - 0.3$ eV. (b) Partial charge density in the energy range of $E_f \sim E_f - 0.3$ eV. (c) Isosurface of differential charge density, $\Delta\rho = \rho(\text{Total}) - \rho(\text{Mo}) - \rho(\text{MgO}) - \rho(\text{OOH}) - \rho(\text{OH})$. Isosurface charge density is taken to be 0.002 electrons/bohr³. Green

(light) and blue (dark) isosurfaces of differential charge density indicate charge accumulation and charge depletion regions. (d) Isosurfaces of spin density difference for spin-up and spin-down states, $\Delta\rho(spin) = \rho(spin\text{-}up) - \rho(spin\text{-}down)$, for $O_2$ and $H_2O$ co-adsorption on 4 ML MgO/Mo(100) surface.

After OOH group is split, the isolated oxygen adatom forms and one surface Mg atom is kicked out from the MgO films. The OH groups show corresponding electronic states between -2 and -4.3 eV, while 2p state from the isolated oxygen adatom is shifted to higher energy level (Figure 4c), suggesting the high reactivity of oxygen adatom. For thicker films, the electrons occupying higher energy level are more than that of thin films (Figure S3), which suggests that isolated oxygen adatom is less favorable on thick film and the metal supported thinner oxide films facilitate the production of high reactive oxygen atom. In addition, the 3σ states of OH split more and more seriously, and the 1π states move to even higher energy levels, which is also responsible for the stable existence of OH adsorbed on very thin oxide films. Due to its high reactivity, oxygen adatom can react with another water to form OH groups without any barrier. After reaction, the electronic states of oxygen adatom are slightly shifted to lower energy region by forming OH group (Figure 4d). All the reactive OH groups are stabilized by the protruded Mg atom. In addition, $O_2$ adsorption also causes the $2p_z$ state of interface O to shift to higher energy region (see Figure S4), indicating the adsorption of $O_2$ has considerable influence in the electronic states at the interface.

In conclusion, we have performed DFT calculations to investigate the co-adsorption of $H_2O$ and $O_2$ on Mo(100) supported ultrathin MgO(100). Interestingly, ultrathin MgO(100) films are dramatically activated by transition metal substrates. Our results indicate the interaction of both $O_2$ and water with Mo(100) supported ultrathin MgO(100) films can stepwise form a series of ROS including hydroperoxide, hydroxyl, superoxide, and single oxygen adatoms. At the beginning of the reaction, the adsorbed water interacts strongly with $O_2$ to form the highly oxidative OOH and OH species without any energy barrier. Then the OOH group can decompose into single oxygen adatom with even more active ability by climbing over the moderate energy barriers on supported ultrathin oxide films. Finally, molecular oxygen completely dissociates with the assistance of water molecule. Our results provide an effective avenue to tune the catalytic activity of supported insulating ultrathin films. More importantly, the generation of ROS with high activities on transi-tion metal supported ultrathin MgO(100) films in a simple way is of scientific and practical importance in chemical reactions.

## ASSOCIATED CONTENT

### Supporting Information

Additional figures. This material is available free of charge via the Internet at http://pubs.acs.org.

## AUTHOR INFORMATION

### Corresponding Author

*xu.h@sustc.edu.cn

### Notes

The authors declare no competing financial interests.

## ACKNOWLEDGMENT

This work is supported by the National Natural Science Foundation of China (NSFC, Grant Nos. 11204185, 11334003 and 11404159) and internal Research Grant Program (FRG-SUSTC1501A-35).

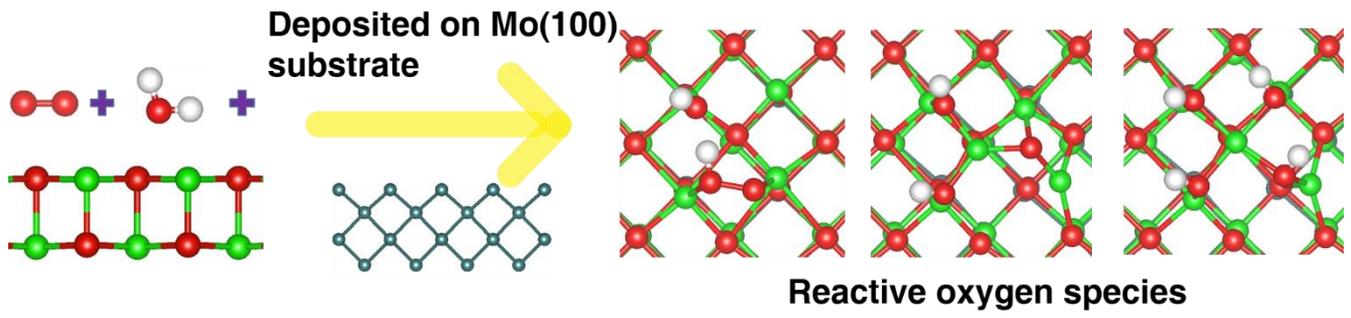

(For Table of Contents Only)